\documentclass[showpacs,amsmath,superscriptaddress,reprint,aps]{revtex4-1}

\usepackage{graphicx}
\usepackage{hyperref}

\begin{document}

\title
{Enhanced Raman scattering of graphene using double resonance in silicon photonic crystal nanocavities}

\author{W.~Gomulya}
\affiliation{Quantum Optoelectronics Research Team, RIKEN Center for Advanced Photonics, Saitama 351-0198, Japan}
\affiliation{Nanoscale Quantum Photonics Laboratory, RIKEN Cluster for Pioneering Research, Saitama 351-0198, Japan}
\author{H.~Machiya}
\affiliation{Nanoscale Quantum Photonics Laboratory, RIKEN Cluster for Pioneering Research, Saitama 351-0198, Japan}
\affiliation{Department of Electrical Engineering, The University of Tokyo, Tokyo 113-8656, Japan}
\author{K.~Kashiwa}
\affiliation{Department of Mechanical Engineering, The University of Tokyo, Tokyo 113-8656, Japan}
\author{T.~Inoue}
\affiliation{Department of Mechanical Engineering, The University of Tokyo, Tokyo 113-8656, Japan}
\author{S.~Chiashi}
\affiliation{Department of Mechanical Engineering, The University of Tokyo, Tokyo 113-8656, Japan}
\author{S.~Maruyama}
\affiliation{Department of Mechanical Engineering, The University of Tokyo, Tokyo 113-8656, Japan}
\affiliation{Energy NanoEngineering Laboratory, National Institute of Advanced Industrial Science and Technology (AIST), Ibaraki 305-8564, Japan}
\author{Y.~K.~Kato}
\email[Corresponding author: ]{yuichiro.kato@riken.jp}
\affiliation{Quantum Optoelectronics Research Team, RIKEN Center for Advanced Photonics, Saitama 351-0198, Japan}
\affiliation{Nanoscale Quantum Photonics Laboratory, RIKEN Cluster for Pioneering Research, Saitama 351-0198, Japan}

\begin{abstract}
We demonstrate enhancements of Raman scattering from graphene on two-dimensional photonic crystals using double resonances, which originate from simultaneous enhancements by a localized guided mode and a cavity mode. By adjusting the photonic crystal cavity parameters, the double resonance can be tuned to the G' Raman scattering. Excitation wavelength dependence measurements show a large Raman peak enhancement when the excitation and emission wavelengths meet the double resonance condition. Furthermore, spatial imaging measurements are performed to confirm that the enhancement is localized at the cavity, and we find that the enhanced Raman intensity is 60 times larger compared to the on-substrate Raman signal. The observed cavity enhancement of Raman scattering opens up new possibilities for the development of graphene-based light sources for silicon photonics.
\end{abstract}

\maketitle

Graphene, a two-dimensional layer of carbon atoms, exhibits remarkable mechanical, electrical, optical, and thermal properties.\cite{Geim:2007} In particular, graphene shows strong light-matter interactions and optical non-linearity, making it an ideal material for optoelectronic devices,\cite{Bao:2012} such as photodetectors,\cite{Xia:2009, Mueller-Graphene:2010} saturable absorbers,\cite{Sun:2010} and optical switches.\cite{Liu:2011} The use of graphene as a light source, however, still remains a challenge due to its gapless nature. While interband radiative recombination is not expected to be strong, graphene exhibits another emission process known as Raman scattering. Remarkably, Raman scattering associated with the G’ mode at $\sim$2700~cm$^{-1}$ in monolayer graphene is stronger than graphite.\cite{Ferrari:2006}

Further emission enhancement by coupling graphene to a nanocavity is desirable for a more efficient light source. Among various cavities, photonic crystal (PhC) cavities hold promise as a tool for increasing the light-matter interactions in nanomaterials by strong electric fields confined in a small mode volume.\cite{Noda:2007} By coupling the emission to the PhC cavities, enhancement of Raman scattering in carbon nanotubes \cite{Sumikura:2013} and quantum dots \cite{Sweeney:2014} have been demonstrated. PhC cavities can also be efficiently coupled to graphene for integrated photonics.\cite{Gan:2012,Gan:2013, Majumdar:2013,Shi:2015,Hwang:2017} Cavity-enhanced Raman scattering from graphene has been achieved by coupling to excitation laser.\cite{Gan:2012}

Here we report on doubly resonant Raman scattering enhancement from monolayer graphene in silicon PhC cavities. The Raman enhancement is obtained by exploiting simultaneous resonance of excitation and emission,\cite{Liu:2015} corresponding to the coupling to a localized guided mode (LGM) and a cavity mode, respectively. Using the double resonance, we are able to enhance the Raman scattering by 60 times compared to that on the un-etched substrate. Our work highlights the potential for using graphene and other two-dimensional materials for monochromatic near-infrared light sources.

\begin{figure}
\includegraphics{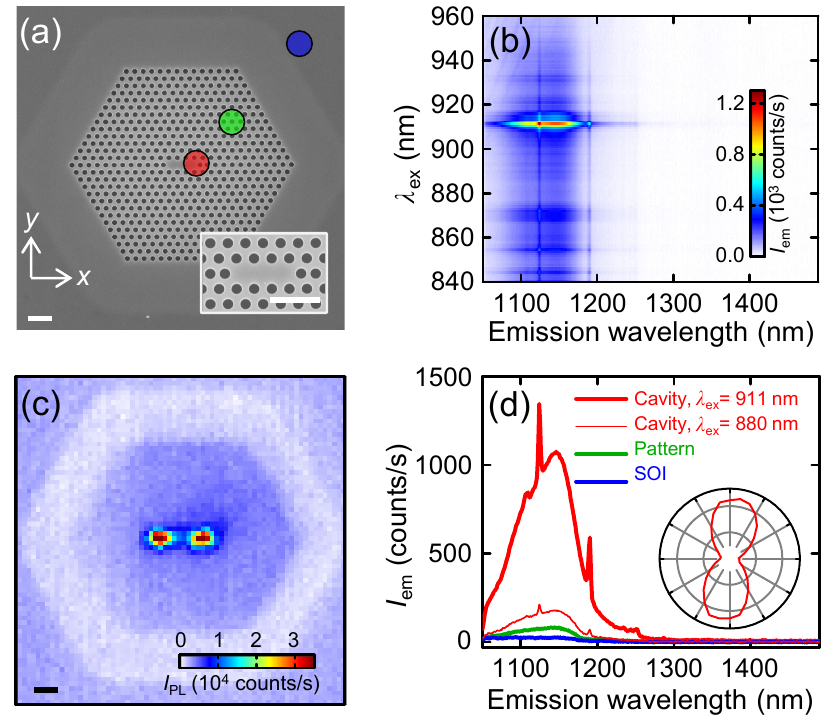}
\caption{
\label{Fig1} (a) An SEM image of a typical device. Red, green, and blue dots indicate the positions at which the PL spectra with corresponding colors in (d) are taken. Inset shows the enlarged view of the cavity. The arrows define the directions of $x$- and $y$-polarization for excitation. The scale bar is 1~$\mu$m. (b) A PL excitation map of a typical PhC cavity. The double resonance is at the intersection between the vertical line corresponding to the cavity mode and the horizontal line corresponding to the LGM. (c) A two-dimensional PL image of the device in (b) measured with $\lambda_{\mathrm{ex}}=911$~nm. The PL intensity $I_{\mathrm{PL}}$ is integrated within a window of 25~nm centered at 1146~nm. The scale bar is 1~$\mu$m. (d) Red, green, and blue thick lines indicate the PL spectra taken at the cavity, patterned area, and un-etched SOI, respectively, excited at $\lambda_{\mathrm{ex}}=911$~nm. The thin red line is a PL spectrum at the cavity excited at $\lambda_{\mathrm{ex}}=880$~nm. Inset shows the excitation polarization dependence of $I_{\mathrm{PL}}$ excited with $\lambda_{\mathrm{ex}}=911$~nm. For (b-d), $y$-polarized laser with a power of 1~mW is used for excitation.}
\end{figure}

The PhC nanocavities are fabricated from silicon-on-insulator (SOI) wafers.\cite{Watahiki:2012, Liu:2015} We define the L3 defect nanocavity by introducing three linear missing air holes in a hexagonal lattice, and we vary the hole radius $r$ and the lattice constant $a$ to tune the LGM and the cavity mode wavelengths.\cite{Fujita:2008, Liu:2015} The positions of the air holes at the ends of the cavity are displaced outward by 0.17~$a$ to improve the quality factor $Q$ of the cavity.\cite{Akahane:2003Nature} The PhC patterns are drawn by electron beam lithography, and the 200-nm-thick top silicon layer is etched through by a dry-etching process. The 1-$\mu$m-thick buried SiO$_{2}$ layer is then removed by 20~wt\% hydrofluoric acid. Figure~\ref{Fig1}(a) shows a scanning electron microscopy (SEM) image of a typical device. The higher magnification image of the cavity clearly shows the shifting of the end holes [Fig.~\ref{Fig1}(a) inset]. 

We characterize the PhC cavities with a home-built laser-scanning confocal microscope. An output of a wavelength-tunable continuous-wave Ti:sapphire laser is focused onto the sample into a spot size of approximately 1 $\mu$m by an infrared objective lens with a numerical aperture of 0.8. The same lens collects the light emission from the samples, and a pinhole corresponding to a 2.7 $\mu$m aperture at the sample imaging plane is placed at the entrance of a 300 mm spectrometer for confocal detection. The emission is recorded by a liquid-nitrogen-cooled InGaAs photodiode array attached to the spectrometer. The sample is placed on a motorized three-dimensional stage to accurately locate the devices, and all measurements are performed at room temperature in a nitrogen environment to prevent graphene oxidation.

We first determine the wavelengths of the LGM and the cavity mode of the PhCs before graphene deposition. The cavities are designed to have double resonances close to the G' Raman condition $\lambda_{\mathrm{ex}}^{-1}-\lambda_{\mathrm{em}}^{-1}=\hbar\omega_{\mathrm{G'}}$, where $\lambda_{\mathrm{ex}}$ is the excitation wavelength, $\lambda_{\mathrm{em}}$ is emission wavelength, $\hbar$ is the Planck constant and $\omega_{\mathrm{G'}}$ is the G' mode frequency. Figure~\ref{Fig1}(b) shows the photoluminescence (PL) excitation map of a typical device with $r = 85$~nm and $a = 355$~nm. The map shows a sharp double resonance originating from the intersection of an LGM at $\lambda_{\mathrm{ex}} = 911$~nm and the 5th mode of the cavity at $\lambda_{\mathrm{em}} = 1190$~nm.

A two-dimensional PL image of the device is measured at $\lambda_{\mathrm{ex}} = 911$~nm with an integration window of 25~nm centered at $\lambda_{\mathrm{em}} = 1146$~nm [Fig.~\ref{Fig1}(c)], showing strong emission localized at the ends of the cavity. Since the emission spectral integration window is chosen not to include any cavity modes, the spatial image reflects the LGM profile. We note that the spatial profile is strongly dependent on the excitation wavelength and excitation polarization.\cite{Liu:2015}

In Fig.~\ref{Fig1}(d), the PL spectra from this device are shown. The thin red line is the PL spectrum on the cavity taken at an off-resonance excitation wavelength $\lambda_{\mathrm{ex}} = 880$~nm. The full-width at half-maximum (FWHM) of the 5th mode of the cavity is 3.9~nm which corresponds to a $Q$-factor of 300. As the excitation wavelength is increased to 911~nm (thick red line), the 5th mode peak emission is enhanced by a factor of 7 compared to that excited at $\lambda_{\mathrm{ex}} = 880$~nm. The increased intensity is due to the excitation field enhancement by the LGM. When we compare the PL at the cavity to the un-etched SOI (blue thick curve), we observe an approximately ninety-fold increase of the cavity peak intensity. Laser excitation polarization dependence measurements at the double resonance show a high degree of excitation polarization [Fig.~\ref{Fig1}(d) inset], where the maximum emission intensity is obtained near $y$-polarized excitation, corresponding to the polarization of the LGM.

After characterization of the silicon PhC cavity, we deposit monolayer graphene flakes on top of the PhC cavities. The monolayer graphene layer is grown on a copper foil by alcohol catalytic chemical vapor deposition (CVD).\cite{Chen:2015} After the growth, the graphene layer is transferred onto the PhCs via wet etching method using poly-methylmethacrylate (PMMA) layer as a mediator. The PMMA layer is spin-coated on top of the graphene on Cu, followed by baking at 100$^\circ$C to solidify the PMMA. Iron trichloride is used to etch the copper foil, and then the graphene attached to the PMMA is placed on the PhCs and left in air overnight to dry. The PMMA layer is then dissolved by immersing the sample in acetone. An additional annealing process at 200$^\circ$C in air for 6 hours is performed to remove any residue.

\begin{figure}
\includegraphics{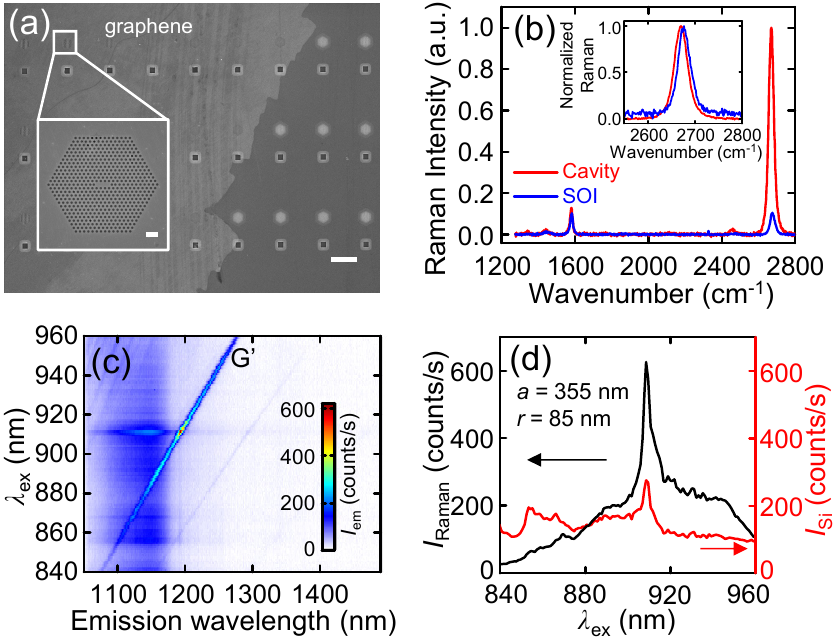}
\caption{ 
\label{Fig2} (a) An SEM image of the devices integrated with CVD-grown graphene. The left side of the image shows devices covered by graphene. Inset shows a higher magnification image of a PhC device. Scale bars in the main image and in the inset are 20~$\mu$m and 1~$\mu$m, respectively. (b) Raman spectra of the monolayer graphene on the cavity (red line) and on the SOI (blue line), taken with 532~nm laser excitation. (c) An excitation spectroscopy map of the G' band Raman scattering of graphene for a device with $a$ = 355 nm and $r$ = 85 nm, taken with a laser power of 1~mW. (d) Raman peak intensity (black) and silicon PL peak intensity (red) obtained by fitting the excitation spectroscopy map in (c).}
\end{figure}

The SEM image of the sample partially covered by graphene is displayed in Fig.~\ref{Fig2}(a), showing high homogeneity of the graphene flake. The inset shows a magnified image of a cavity. To verify the layer number, Raman spectroscopy under 532~nm laser excitation is performed at different points on the sample. Typical on-cavity (red line) and on-SOI (blue line) Raman spectra are shown in Fig.~\ref{Fig2}(b). On the cavity, two sharp peaks centered at 1580~cm$^{-1}$ and 2670~cm$^{-1}$ corresponding to the G and G' bands, respectively, are observed. The ratio of the G' peak intensity $I_{\mathrm{G'}}$ to G peak intensity $I_{\mathrm{G}}$ of 8 for the on-cavity graphene confirms that the graphene is a single layer.\cite{Ferrari:2006} The G' band also shows symmetric Lorentzian profile with a FWHM of 30.6~cm$^{-1}$, giving another evidence for monolayer graphene.\cite{Wang:2008, Graf:2007} The on-SOI G' Raman peak has $I_{\mathrm{G'}}/I_{\mathrm{G}} = 0.96$, and its center wavelength is blueshifted by 7~cm$^{-1}$ compared to the on-cavity Raman peak. Similar blueshift and weaker G' intensity have been reported in the case of Raman scattering of graphene on SiO$_2$/Si substrate to be due to unintentional doping from the interaction of the graphene with the substrate.\cite{Berciaud:2009}

Figure~\ref{Fig2}(c) shows the excitation spectroscopy map of the device in Fig.~\ref{Fig1}(b) with graphene on top, where the cavity mode intensity at 1190~nm has become lower after the graphene deposition. The bright diagonal line on the excitation spectroscopy map arises from the G' Raman scattering of the graphene. The Raman intensity increases when the G' mode is tuned to the double resonance, and then decreases as the excitation wavelength is detuned from the double resonance.

To quantitatively calculate the Raman enhancement, we perform curve fitting of emission spectra for each excitation wavelength. Because the complex silicon PL spectra cannot be fitted using a simple function, we use a linear combination of an empirical spectrum of silicon PL and a Lorentzian peak
\begin{equation}
I_{\mathrm{em}}(\lambda_{\mathrm{em}})=I_{\mathrm{Si}}S(\lambda_{\mathrm{em}})+I_{\mathrm{Raman}}\frac{(w/2)^{2}}{(\lambda_{\mathrm{em}}-\lambda_{\mathrm{0}})^{2}+(w/2)^{2}}+I_{\mathrm{0}},
\end{equation}
where $I_{\mathrm{em}}(\lambda_{\mathrm{em}})$ is the emission spectrum, $I_{\mathrm{Si}}$ is the silicon PL peak intensity, $S(\lambda_{\mathrm{em}})$ is the normalized silicon spectrum, $I_{\mathrm{Raman}}$ is the Raman peak intensity, $\lambda_{\mathrm{0}}$ is the center wavelength of the Raman peak, $w$ is the Raman peak FWHM, and $I_{\mathrm{0}}$ is the offset intensity. In Fig.~\ref{Fig2}(d), $I_{\mathrm{Raman}}$ and $I_{\mathrm{Si}}$ are plotted as a function of the excitation wavelength. To quantify the resonant enhancement, we use the ratio between the on-resonance Raman peak and the off-resonance Raman peak at $\lambda_{\mathrm{ex}} = 960$~nm. The calculated ratio for the device shown in Fig.~\ref{Fig2}(c) is 5.9. We have performed similar measurements and analyses on 37 devices with the same design parameters, and we obtain an average enhancement ratio of 6.1 with a standard deviation of 1.0 and a maximum enhancement ratio of 8.5.

\begin{figure}
\includegraphics{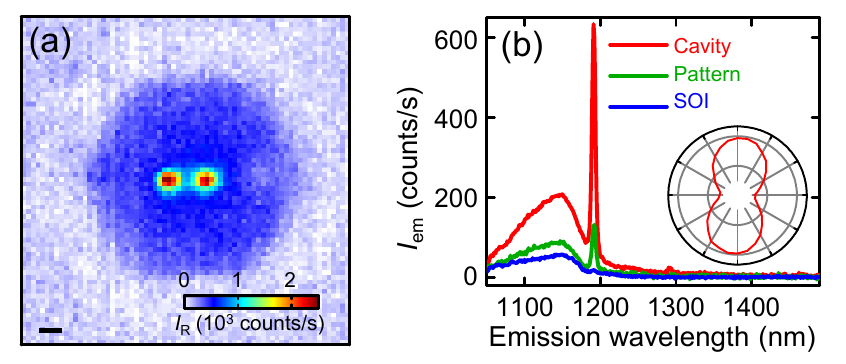}
\caption{
\label{Fig3}(a) A two-dimensional emission image of the device in Fig.~2(c). The emission intensity $I_{\mathrm{R}}$ is integrated over a spectral window centered at the Raman peak wavelength of 1190~nm with a width of 5.2~nm. The scale bar is 1~$\mu$m. (b) Emission spectra of the graphene on PhC taken at the cavity (red), on the pattern (green), and on the un-etched SOI (blue). Inset shows the excitation polarization dependence of $I_{\mathrm{Raman}}$ at the double resonance. For (a-b),  $y$-polarized laser with $\lambda_{\mathrm{ex}}=911$~nm and a power of 1~mW is used for excitation.}
\end{figure}

To confirm that the enhancement originates from the Raman emission coupled to the double resonance of the PhC cavity, we explore the spatial extent of the Raman signal. The two-dimensional image of the Raman peak measured with $\lambda_{\mathrm{ex}} = 911$~nm is displayed in Fig.~\ref{Fig3}(a), and we observe a spatial profile similar to the silicon PL image before graphene deposition [Fig.~\ref{Fig1}(c)]. On this device, the highest Raman scattering intensity is found at the ends of the cavity. We also find that the Raman intensity on the PhC pattern is higher than the Raman scattering intensity on the SOI, which can be explained by the higher collection efficiency of the emitted light in the direction normal to the slab plane through the leaky modes.\cite{Fujita:2008, Fujita:2005, Fan:2002}

Figure~\ref{Fig3}(b) shows typical emission spectra of the graphene at different positions of the PhC structure with the same color convention as in Figs.~\ref{Fig1}(a) and \ref{Fig1}(d), where the red curve shows the strongest Raman intensity. Although the cavity emission due to the silicon PL is barely visible for off-resonant excitation, we calculate the enhancement conservatively by assuming that the on-cavity emission peak contains signals from Raman scattering and silicon PL. We estimate the contribution of the silicon PL from the spectrum taken on the device before graphene deposition [Fig.~\ref{Fig1}(d)], and find that the double-resonance-enhanced cavity peak at 1190~nm is 40\% of the broad silicon peak intensity at 1146~nm. The on-cavity Raman peak value taken from the red curve in Fig.~\ref{Fig3}(b) is then corrected by subtracting the estimated silicon contribution using this fraction. After the correction, we obtain a value of 4.2 for the enhancement of the Raman scattering by the PhC cavity compared to the Raman signal on the PhC pattern (green curve). By comparing the on-cavity Raman signal to that on the un-etched SOI (blue curve), the PhC cavity gives an enhancement by a factor of 60. 

Furthermore, we perform polarization dependence measurements [Fig.~\ref{Fig3}(b) inset], and the Raman peak is found to have a similar excitation polarization profile as the LGM. The maximum intensity of the Raman scattering is 3.3 times the minimum Raman intensity which would correspond to the Raman intensity without LGM enhancement. This result is similar to a previous report \cite{Gan:2012} showing absorption enhancement factor of 3.41 for a PhC cavity with a $Q$-factor of 330.

\begin{figure}
\includegraphics{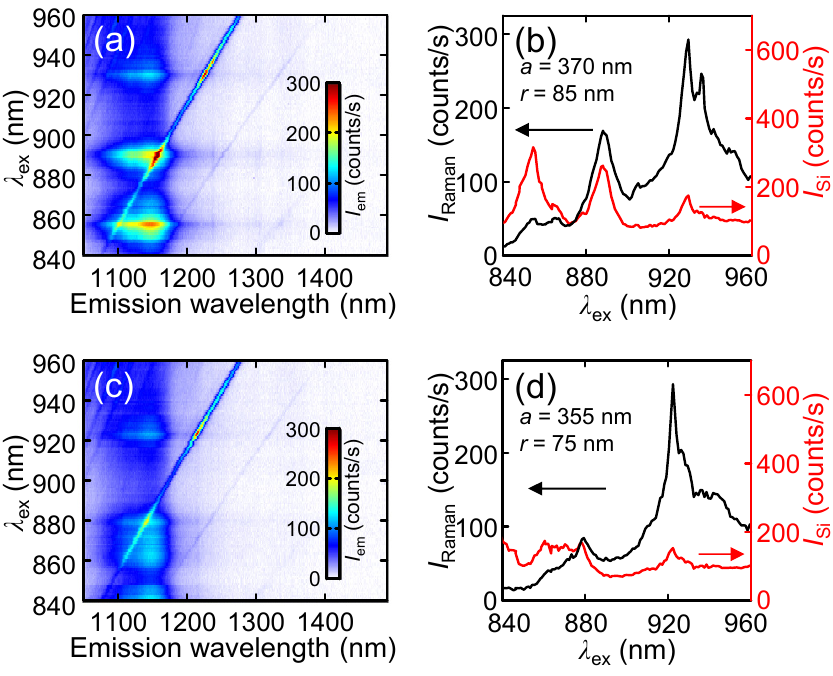}
\caption{
\label{Fig4} Excitation spectroscopy maps taken with a laser power of 1~mW for (a) a device with a double resonance not located on the Raman line and (c) a device with a double resonance on the Raman line with longer wavelength compared to the device in Fig.~\ref{Fig2}(c). (b) and (d) Excitation wavelength dependence of Raman peak intensity (black) and silicon PL peak intensity (red) of the devices in (a) and (c), respectively.}
\end{figure}

By investigating a device with a different lattice constant, it is possible to separate the enhancement by the cavity mode and the LGM. One typical excitation spectroscopy map for a device with $r = 85$~nm and $a = 370$~nm is shown in Fig.~\ref{Fig4}(a), where the double resonance shifts to the right side of the diagonal Raman line. By performing a similar analysis as before, we obtain fitting results of this device [Fig.~\ref{Fig4}(b)]; the black curve for  $I_{\mathrm{Raman}}$ and the red curve for $I_{\mathrm{Si}}$. The excitation spectroscopy map and the Raman excitation spectra in Fig.~\ref{Fig4}(a) and \ref{Fig4}(b) clearly show two peaks at excitation wavelengths of 930~nm and 937~nm, corresponding to Raman scattering enhanced by the LGM and the cavity mode, respectively. Again, using the ratio with respect to the off-resonance Raman peak at $\lambda_{\mathrm{ex}}=960$~nm, the Raman enhancement by the LGM is calculated to be 2.7. The enhancement by the cavity mode is calculated differently due to the spectral overlap with the LGM (supplementary material Fig.~S1). After excluding the enhancement from the LGM, we find an enhancement of 1.3 by the cavity mode.

Finally, we demonstrate wavelength tuning of the Raman enhancement by the double resonance. In contrast to varying the lattice constant, changing the PhC air hole radius moves the double resonance nearly along the Raman line. One typical excitation spectroscopy map for a device with $r = 75$~nm and $a = 355$~nm is shown in Fig.~\ref{Fig4}(c) and its fitting results are displayed in Fig.~\ref{Fig4}(d). The double resonance is located at $\lambda_{\mathrm{ex}} = 924$~nm and $\lambda_{\mathrm{em}} = 1214$~nm, redshifted compared to that for the device shown in Figs.~\ref{Fig2}(c) and \ref{Fig2}(d). The enhancement by double resonance for this device is 2.8 when compared to the Raman peak at $\lambda_{\mathrm{ex}}=960$~nm. Similar measurements are performed on devices with different $r$, and we find that as the hole radius decreases, the double resonance wavelengths become longer and the enhancement becomes lower (See supplementary material Fig.~S2).

To further improve the Raman scattering emission, optimization of the quality factor of our devices is required. The LGM has lower quality factor which could potentially be increased by engineering the PhC structure.\cite{Fan:2002, Tandaechanurat:2008} Once we obtain a higher enhancement factor, light emission can be made stronger by improving Raman scattering intensity, which could be achieved, for example, by surface-enhanced Raman scattering.\cite{Schedin:2010}

In summary, we have used double resonances in PhC cavities to enhance the Raman scattering of graphene. We have successfully designed the PhC nanocavities to have double resonances on the G' Raman line, and we obtain the maximum Raman intensity when the laser is tuned to meet the resonant excitation and emission conditions. Spatial imaging measurements confirm that the enhancement originates from the coupling to the cavity double resonance, and we observe an enhancement of the Raman intensity by a factor of 60 compared to that on the un-etched SOI. By varying the device lattice constant, we are able to separate and estimate the enhancement contributions from the LGM and the cavity mode. Furthermore, the enhanced Raman emission wavelength is tunable by varying the hole radius of the PhC. Our results mark an important step towards the development of monochromatic near-infrared light sources using graphene and other two-dimensional materials for integrated silicon nanophotonics, and may provide a way to study electron-phonon interaction in nanomaterials.

\begin{acknowledgments}
Work supported by JSPS (KAKENHI JP16K13613, JP25107002) and MEXT (Photon Frontier Network Program, Nanotechnology Platform). W.G. is an International Research Fellow of JSPS (Postdoctoral Fellowship for Research in Japan (Standard)). H.M. is supported by RIKEN Junior Research Associate Program. We acknowledge technical support from Advanced Manufacturing Support Team, RIKEN.
\end{acknowledgments}

%\bibliographystyle{pr}
%\bibliography{References}
%merlin.mbs apsrev4-1.bst 2010-07-25 4.21a (PWD, AO, DPC) hacked
%Control: key (0)
%Control: author (72) initials jnrlst
%Control: editor formatted (1) identically to author
%Control: production of article title (-1) disabled
%Control: page (0) single
%Control: year (1) truncated
%Control: production of eprint (0) enabled
%

\end{document}